\documentclass[12pt]{article}
\pdfoutput=1
\usepackage[body={17.5cm, 22cm},right=2cm]{geometry}
\usepackage{amssymb,graphicx}
\usepackage{amsmath}
\usepackage{epsfig}
\usepackage{hyperref}

\newcommand{\be}{\begin{equation}}
\newcommand{\ee}{\end{equation}}
\newcommand{\bea}{\begin{eqnarray}}
\newcommand{\eea}{\end{eqnarray}}
\newcommand{\bg}{\begin{gather}}
\newcommand{\eg}{\end{gather}}
\newcommand{\bseq}{\begin{subequations}}
\newcommand{\eseq}{\end{subequations}}

\begin{document}
\setcounter{page}{0}
\thispagestyle{empty}

\parskip 3pt

\begin{titlepage}
%\begin{flushright}
% SLAC-PUB-1XXXX
%\end{flushright}
%
~\vspace{1cm}
\begin{center}

{\LARGE \bf A Non Standard Model Higgs at the LHC \\[.4cm] as a Sign of Naturalness}

\vspace{1.2cm}

{\large \bf
Asimina Arvanitaki$^{a}$,
Giovanni Villadoro$^{b}$}
\\
\vspace{.6cm}
{\normalsize { \sl $^{a}$ Stanford Institute for Theoretical Physics,\\
Stanford University, Stanford, CA 94305 USA}}

\vspace{.3cm}
{\normalsize { \sl $^{b}$ SLAC, Stanford University \\
2575 Sand Hill Rd., Menlo Park, CA 94025 USA}}

\end{center}
\vspace{.8cm}
\begin{abstract}
Light states associated with the hierarchy problem affect the Higgs LHC production and decays.
We illustrate this within the MSSM and two simple extensions applying the latest bounds from LHC Higgs searches. 
Large deviations in the Higgs properties are expected in a natural SUSY spectrum.
The discovery of a non-Standard-Model Higgs may signal the presence of light stops accessible
at the LHC. Conversely, the more the Higgs is Standard-Model-like, the more tuned the theory becomes.
Taking the ratio of different Higgs decay channels at the LHC cancels the leading QCD uncertainties and  potentially improves the accuracy in Higgs coupling measurements to the percent level. This may lead to the possibility of doing precision Higgs physics at the LHC.
Finally, we entertain the possibility that the ATLAS excess around 125 GeV persists with a Higgs production cross-section that is enhanced compared to the SM. This increase can only be accommodated in extensions of the MSSM and it may suggest that stops lie below 400 GeV, likely within reach of next year's LHC run.
\end{abstract}

\end{titlepage}

\tableofcontents

\section{Introduction}
The Higgs boson is the last missing ingredient of the Standard Model (SM) and it has been
the focus of collider searches in the past 30 years. It explains electroweak (EW) symmetry breaking 
and the particle masses but at the price of a huge fine-tuning needed to keep the scale
of weak interactions far below the Planck scale. Any solution to this hierarchy problem requires
new degrees of freedom at the EW scale to cancel the quadratic divergences of the Higgs:
\begin{center}
\includegraphics[width=10cm]{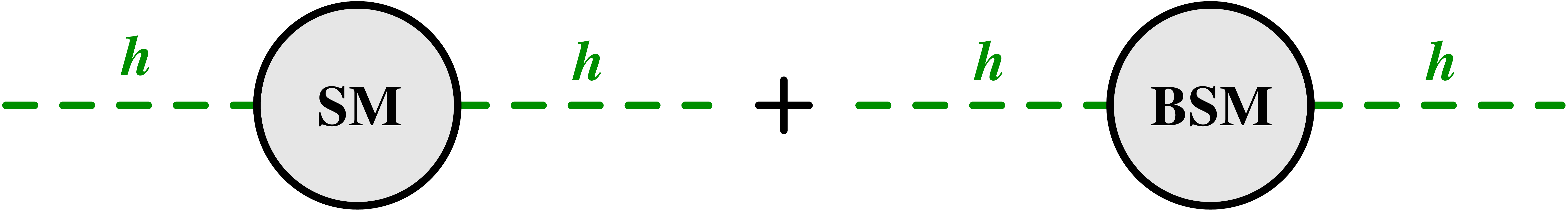}
\end{center}
These degrees of freedom  also affect the main Higgs production mechanism at the Large Hadron Collider (LHC), gluon-gluon fusion:
%At the large hadron collider (LHC) the main Higgs production mechanism, the gluon-gluon fusion,
%is controlled by the dof that are mainly responsible for the hierarchy problem and its solution:
\begin{center}
\includegraphics[width=10cm]{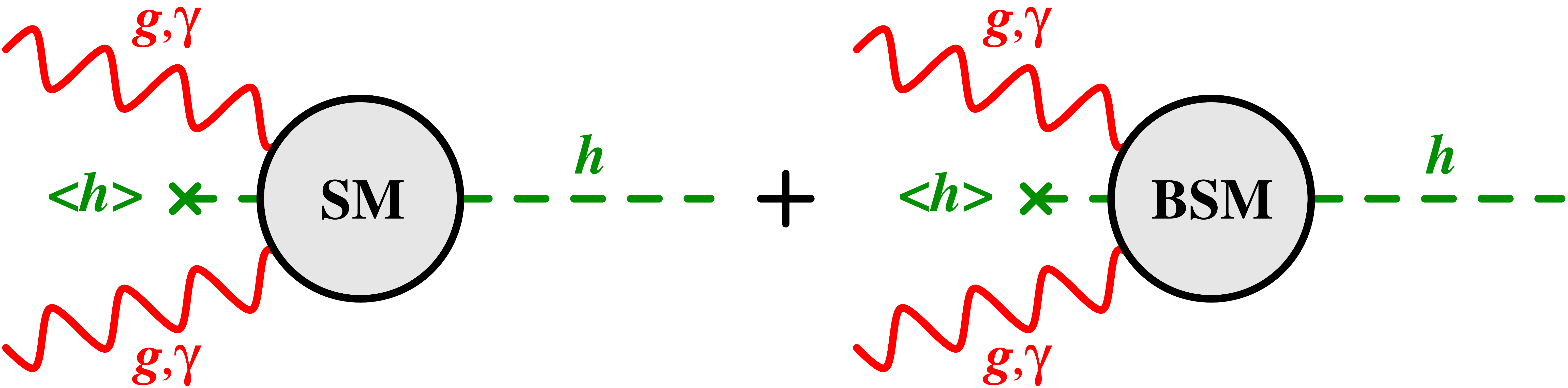}
\end{center}
The same is true for one of the most important Higgs decay channels, $h\to \gamma\gamma$.
Therefore we expect a natural Higgs at the LHC to behave differently from the SM Higgs.
Vice-versa the more the Higgs is SM-like the more tuned the theory will be.
This fact highlights the importance of accurately measuring the Higgs couplings to SM particles.

In this paper we study the interplay between naturalness and Higgs properties at the LHC
in the framework of supersymmetry (SUSY) \cite{Dimopoulos:1981zb} . Supersymmetry is still the leading candidate
for physics beyond the SM, it predicts gauge coupling unification, and provides a viable 
dark matter candidate.
% at the EW scale with calculable abundance.
How natural is SUSY, though, given the absence of any significant deviation from the SM?
Within the minimal implementation of SUSY (MSSM), the LEP bounds on the Higgs mass
increase the fine-tuning to an uncomfortable level, but this problem is solved in simple
extensions of the MSSM. Still, current LHC bounds on squark masses naively push
the scale of SUSY around 1~TeV \cite{Aad:2011ib, Chatrchyan:2011zy} and reintroduce the fine-tuning problem. These
bounds, however, apply for universal squark masses, while naturalness in SUSY requires
only the top squarks to be light~\cite{Dimopoulos:1995mi}. Direct searches for third family sparticles
allow for stops as light as the top and SUSY can be natural~\cite{Shih, Papucci:2011wy}. In this case
light stops can significantly alter LHC Higgs production and decay.

Besides the stop sector, the tree level Higgs couplings to SM particles also depend 
on the mixing of the light Higgs with the other Higgses in SUSY. 
The Higgs behavior at the LHC and the tuning can be adequately described by just
three parameters: the lightest stop mass, the stop mixing and the Higgs mixing angle.
We study how the Higgs production and decay increase or decrease with respect to the SM
as a function of these three parameters in the MSSM and two simple extensions.
We also discuss how taking the ratio of different decay channels 
improves the accuracy for Higgs coupling measurements at the LHC.
Finally, we comment on the recent ATLAS excess at $\sim$125~GeV and, if confirmed, what
this would imply for the SUSY spectrum.

\section{The Higgs in Supersymmetric Models}
\label{SUSYHiggs}

The light Higgs mass in the MSSM is naturally constrained to lie below the $Z$ mass at tree level. The LEP bounds~\cite{PDG} put tremendous strain on SUSY where large radiative corrections are required to raise the Higgs mass. This also increases the amount of tuning that is needed to account for the measured $G_F$. The amount of tuning can be quantified by the relation~\cite{Barbieri:1987fn}:
\bea
FT(\xi)=\left [\sum_{i} \left(\frac{\partial \log v^2}{\partial \log \xi^i}\right)^2\right]^{-1/2}\,,
\label{fine-tuning}
\eea
where $\xi_i$ includes all the parameters that contribute to $v$. 
In the MSSM, the $Z$ mass is, in the large $\tan \beta$ limit:
\bea
m_Z^2\approx-2(m_{H_u}^2+\mu^2)+\frac{2}{\tan^2\beta}(m_{H_d}^2-m_{H_u}^2)\,.
\eea

As can be seen from this formula, the main contributors to the tuning are the $\mu$-term and  $m_{H_u}$. 
$m_{H_u}$ also receive large radiative corrections,
\begin{equation}
\delta m_{H_u}^2\approx \frac{3 y_t^2}{16\pi^2}\left(m_{\tilde t_1}^2+m_{\tilde t_2}^2+A_t^2\right)
\log\left(\frac{2\Lambda^2}{m_{\tilde t_1}^2+m_{\tilde t_2}^2}\right)\,,
\end{equation}
from the stop masses, $m_{\tilde t_{1,2}}$, and mixing, $A_t$, which are also the parameters 
responsible for raising the Higgs mass. 
As a result, naturalness (say $FT\gtrsim10\%$) pushes Higgsinos and stops to be light:
\bea
\mu \lesssim 250~\text{GeV, and~}m_{\tilde t}\lesssim 350~\text{GeV}\,.
\eea
It is worth noting that $m_{H_d}$ does not affect the tuning significantly at the large $\tan \beta$ limit, so the rest of the Higgs sector in the MSSM can be above 1 TeV without loss of naturalness.

The LHC production of the Higgs is primarily modified by the same particles that contribute the most to the radiative corrections of its mass and to the tuning: the stops. Even though current LHC searches exclude the possibility of universal squark spectra up to almost 1~TeV ($\sim$500~GeV with a heavy neutralino), stops can be perhaps as light as the top in models with non-universal squark masses~\cite{Dimopoulos:1995mi} or when SUSY is hidden, for example  see~\cite{Fan:2011yu}. 
As a result, stops can significantly influence the Higgs coupling to gluons. 

In the supersymmetric limit, light stops increase the LHC Higgs production cross-section~\cite{Gunion:1984yn, Gunion:1986nh, Kileng, Spira:1995rr, Djouadi:1998az}, since the latter is related to the $\beta$-function of $\alpha_s$ to which scalar fields contribute with the same sign as fermions~\cite{Shifman:1979eb, Low:2009di}. This behavior persists after SUSY breaking in the absence of stop mixing, while a large mixing between the stop eigenstates interferes destructively with the top loop and the gluon fusion process can be suppressed substantially. Neglecting bottom/sbottom and D-term contributions, the approximate formula for the $gg\to h$ production cross-section of a light Higgs ($m_H \lesssim m_t $) relative to the Standard Model is
\bea
\frac{\sigma(gg\to h)}{\sigma_{SM}(gg\to h)}\approx 
\left[ 1+\frac14 \left( \frac{m^2_t}{m^2_{\tilde t_1}}+\frac{m^2_t}{m^2_{\tilde t_2}}-\frac{m_t^2 (A_t+\mu/\tan\beta)(A_t-\mu\tan\alpha)}{m^2_{\tilde t_1}m^2_{\tilde t_2}} \right)\right]^2\,,
\eea    
where $m_{\tilde t_i}$ are the stop mass eigenvalues and $\alpha$ is the mixing angle between the up and down
type Higgses. This effect can be significant even for stops as heavy as 400~GeV ($\sim 20\%$ in the absence of mixing). Notice that the mixing term is always negative since $(A_t-\mu\tan\alpha)\simeq(A_t+\mu/\tan\beta)$, which is a good approximation except in the extreme case of large mixing between the light and the heavy Higgs. 

For Higgs masses below $\sim$135~GeV, the branching ratio (BR) to photons is also important. Similarly to gluon fusion, the Higgs coupling to two photons is again related to the electroweak $\beta$-functions. In this case, the dominant contribution comes from the $W$ boson while the top gives a smaller contribution with opposite sign. With large $A$-terms, the stop contributions, being opposite to the top, increase the Higgs BR to photons slightly, but this enhancement is smaller than the suppression from the production cross section due to the minor role of the top in the diphoton channel. Chargino states also contribute to $h\rightarrow \gamma \gamma$, but once they become heavier than roughly 100~GeV their effects are subleading.

The other important deviation of the SUSY Higgs' behavior from the SM comes from its natural embedding in two Higgs doublets. The role of the SM Higgs is shared by two Higgs states, $H_u$ and $H_d$, and the mixing angle $\alpha$ between $H_u$ and $H_d$ now affects all the tree level couplings of the light scalar Higgs state. The parameters that mostly influence the amount of mixing is the pseudoscalar Higgs mass $m_A$ and $\tan\beta$. In the limit of large $\tan\beta$ 
the shift of the top, bottom and vector boson couplings of the light Higgs compared to their SM values are
\begin{align}
\label{yukawasMSSM}
\frac{\delta y_t}{y_t}&=\frac{\cos\alpha}{\sin\beta}-1\approx-  \frac{1}{\tan^2\beta} \frac{2\,\frac{m_Z^2}{m_A^2}}{\left(1-\frac{m_Z^2}{m_A^2}\right)^2}
%\frac{\delta y_t}{y_t}&=\frac{\cos\alpha}{\sin\beta}-1\stackrel {\tan\beta\to\infty} {\xrightarrow{\hspace*{1.2cm}}}-  \frac{1}{\tan^2\beta} \frac{2\,\frac{m_Z^2}{m_A^2}}{\left(1-\frac{m_Z^2}{m_A^2}\right)^2}
%\stackrel {m_A\gg m_Z} {\xrightarrow{\hspace*{1.2cm}}}- 2\frac{m_Z^2}{m_A^2} \frac{1}{\tan^2\beta} 
\,, \\
\frac{\delta y_b}{y_b}&=-\frac{\sin\alpha}{\cos\beta}-1\approx \frac{2\,\frac{m_Z^2}{m_A^2}}{1-\frac{m_Z^2}{m_A^2}} 
%\frac{\delta y_b}{y_b}&=-\frac{\sin\alpha}{\cos\beta}-1\stackrel {\tan\beta\to\infty} {\xrightarrow{\hspace*{1.2cm}}} \frac{2\,\frac{m_Z^2}{m_A^2}}{1-\frac{m_Z^2}{m_A^2}} %\stackrel {m_A\gg m_Z} {\xrightarrow{\hspace*{1.2cm}}}2\frac{m_Z^2}{m_A^2} 
\,,\\
\frac{\delta g_{VV}}{g_{VV}}&= \sin(\beta-\alpha)-1\approx -\frac{1}{\tan^2\beta}\frac{2\,\frac{m_Z^4}{m_A^4}}{\left(1-\frac{m_Z^2}{m_A^2}\right)^2} 
%\frac{\delta g_{VV}}{g_{VV}}&= \sin(\beta-\alpha)-1\stackrel {\tan\beta\to\infty} {\xrightarrow{\hspace*{1.2cm}}}  \frac{1}{\tan^2\beta}\frac{2\,\frac{m_Z^4}{m_A^4}}{\left(1-\frac{m_Z^2}{m_A^2}\right)^2} 
%\stackrel {m_A\gg m_Z} {\xrightarrow{\hspace*{1.2cm}}} - 2 \frac{m_Z^4}{m_A^4}\frac{1}{\tan^2\beta}
\,,
\end{align}
respectively. This expansion shows that mixing becomes important for a light Higgs between 115 and $\sim$150 GeV, when the BR to $b \bar{b}$ is dominant. Mixing in this case can suppress the Higgs cross-section in a given channel compared to the SM value significantly (more than 20$\%$) when $m_A$ is lighter than roughly 400 GeV. 
A special region in two Higgs doublet models is the region of maximal mixing, where the role of the SM is almost equally shared between the light and the heavy Higgs, which are now close in mass. The effects of this region will be examined separately in section~\ref{Higgsmixing}.

From the above discussion, it is obvious that the properties of the SUSY Higgs can be adequately described by three parameters: the lightest stop mass $m_{\tilde t_2}$, the SUSY breaking $A$-term in the stop sector $A_t$, and the pseudoscalar Higgs mass $m_A$\footnote{We will not vary the sbottom sector contributions as their effects are only important at extremely large values of $\tan\beta$ and maximal Higgs mixing.}. 
Another parameter that affects the LHC search strategy for the Higgs is its mass. When the Higgs is between 115 and $\sim$135 GeV, the most sensitive channel at the LHC is the Higgs decay to two photons. Above those masses, Higgs decay to $WW^{(*)}$ becomes most important and completely dominates the Higgs BR above  $2 m_{W}\approx 160$ GeV. At the same time, the Higgs decay to $ZZ^{(*)}$ increases the LHC sensitivity and becomes the main search channel for  Higgs heavier than roughly 200 GeV. In what follows we will consider the latest results in Higgs searches~\cite{ATLAShiggsresults, CMShiggsresults}, which are most sensitive in $h\rightarrow \gamma \gamma$ when the Higgs is lighter than 131.5 GeV, in $h\rightarrow W W^{(*)}$ for $131.5~\text{GeV}<m_h<2 m_W$, and the combined $ZZ^{(*)}$ and $WW$ above the WW threshold. 

It is well known that, after LEP and Tevatron bounds on the Higgs mass and the sparticle spectrum, within the MSSM the fine-tuning of the parameters is never better than few percent (this is achieved in the golden region of large $A$-terms \cite{perelstein}). The situation improves somewhat in the presence of extra tree-level contributions to the Higgs mass, which relax the tuning of the weak scale. For this reason, in the rest of the paper, we will consider, besides the case
of an MSSM Higgs, two other cases: 1) MSSM with extra D-terms (DMSSM), where an extra quartic term for the Higgs is generated by an additional gauge sector that is broken above the electroweak scale~\cite{D-terms}, and 2) NMSSM~\cite{NMSSMreview}($\lambda SUSY$~\cite{Harnik:2003rs, lambdaSUSY}), where the extra contribution to the Higgs mass comes from an extra singlet (with possibly large couplings). 

%In the following sections we will compare the SUSY Higgs properties to those of its SM counterpart as a function of $m_A$, $A_t$  and $m_{\tilde t_2}$ in three scenarios that cover a large part of the Higgs mass range accessible to the LHC: 
%\begin{itemize}
%\item{The Minimal Supersymmetric Standard Model}\\
%\item {The case of non-decoupling D-terms for the Higgs}\\
%\item{The Next-to-Minimal Supersymmetric Standard Model}\\
%\end{itemize}

\subsection{Collider Bounds}
\begin{figure}[t] %  figure placement: here, top, bottom, or page
 \begin{center}
 \includegraphics[height=8cm]{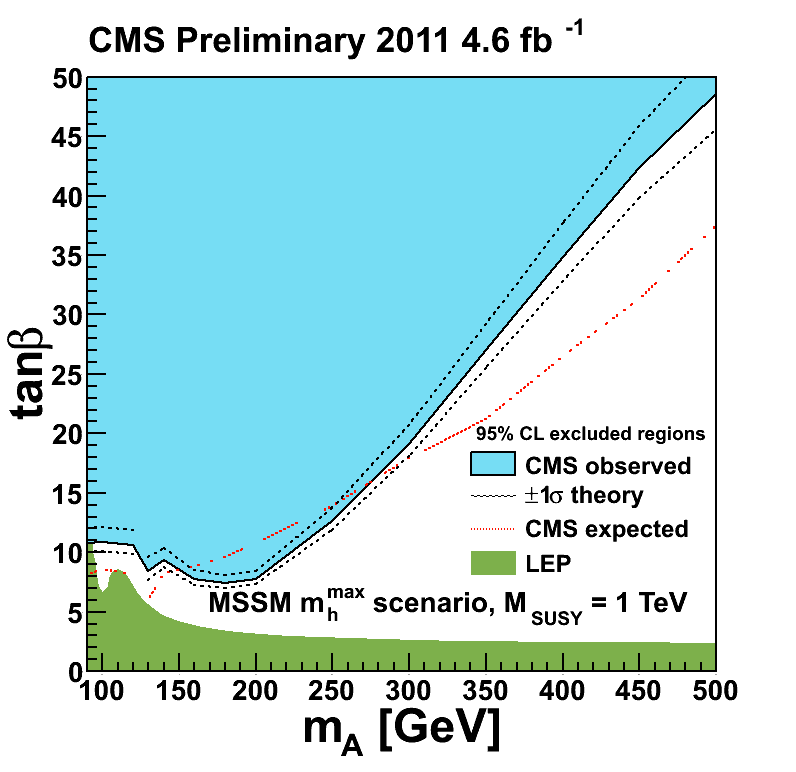}
 \caption{CMS bounds to $m_A$ as a function of $\tan\beta$ from~\cite{mAbounds}.}
 \label{fig:tanbma}
 \end{center}
\end{figure}

So far collider searches have focused mainly on universal sparticle spectra with LHC placing bounds that range from 800 to 1000 GeV for squarks masses\cite{Aad:2011ib, Chatrchyan:2011zy}. It is worth pointing out that these bounds are relaxed to roughly 500 GeV when the LSP mass is 200 GeV lighter than the squarks~\cite{Chatrchyan:2011zy}. Still, there are very few searches that focus on the possibility of only a light 3rd family of sparticles. Given the reduced production cross section in this case the bounds are weaker and split family spectra allow for less tuning in SUSY.

At the LHC, 3rd sparticle family searches primarily rely on scalar production from gluino decays. Since the gluino contribution to the Higgs mass enters at two loops, they can be heavier than TeV and above the present LHC reach. A reanalysis of LHC results \cite{Papucci:2011wy} suggests that bounds lie around 200-300~GeV depending on whether one or two stops are light but may disappear if the LSP is a pure bino, see also \cite{Shih}. The bounds on the stops can also be relaxed in various scenarios where SUSY is hidden. In our analysis we will thus allow $m_{\tilde t_2}$ to vary from the top mass to roughly 800~GeV as suggested by naturalness.

For the SUSY Higgs sector the biggest constraint comes from the LEP and the LHC searches in the $b \bar b$  and $\tau^+\tau^-$ channel respectively~\cite{mAbounds}. In most of the SUSY parameter space, where there is no large mixing between the light and the heavy Higgs, the light scalar Higgs is constrained to be above $\sim$114~GeV. This is the main source of  fine-tuning in the MSSM as it naively pushes stop masses above 1~TeV. LEP and LHC bounds on the light Higgs also place a bound on $m_A$ of roughly 100~GeV (see Fig.~\ref{fig:tanbma}). 

Charged Higgs searches are not as constraining with LEP bounds on the charged Higgs around $\sim$75 GeV and searches for $t \rightarrow H^+ b$ at hadron colliders not very sensitive yet~\cite{PDG}.

\subsection{Bounds from $b\to s \gamma$}

The strongest constraints on a light stop-Higgs sector come from indirect bounds from rare $B$ decays, in particular  for the moderate values of $\tan\beta$ considered in this work the constraints are dominated by $b\to s\gamma$. The main contributions come from the charged Higgs-top and the chargino-stop loops. The charged Higgs increases the amplitude of $b\rightarrow s  \gamma$ and varies mildly with $\tan\beta$ unless $\tan\beta\lesssim1$. When all other SUSY particles are decoupled, the charged Higgs is constrained to be heavier than $\sim$350~GeV. 

\begin{figure}[t!] %  figure placement: here, top, bottom, or page
 \begin{center}
 \includegraphics[width=7cm]{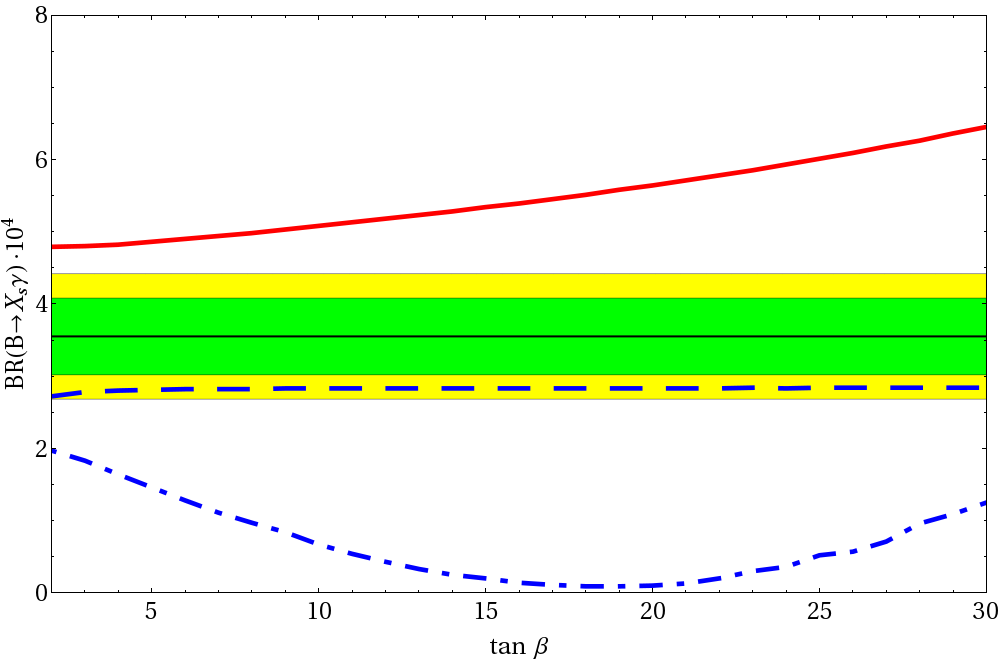}
  \caption{$BR(\bar B\to X_s\gamma)$~$vs$~$\tan\beta$ plot. The green-yellow band correspond to the 1$\sigma$-2$\sigma$
  experimental bound.   All the SUSY particles have been decoupled except 1) $m_A=250$~GeV for the red line, 2) $\mu=m_{\tilde t_L}=m_{\tilde t_R}=200$~GeV, $A_t=0$ for the dashed blue line, 3) $\mu=m_{\tilde t_L}=m_{\tilde t_R}=A_t=200$~GeV for the dot-dashed blue line.}
 \label{fig:bsg-tanb}
 \end{center}
\end{figure}

This bound changes drastically when charginos and stops are light~\cite{bsg-barbierigiudice}. Chargino contributions interfere destructively or constructively with the SM depending on the relative sign of the $\mu$ and $A$-terms. They significantly affect $b\to s \gamma$ at large $\tan\beta$. Their effect is minimized when the rest of the squarks are also light and the $A$-term are small due to the GIM mechanism. A natural spectrum requires light Higgsinos and stops,
and large cancellations from a light charged Higgs or from $A$-terms are necessary to satisfy the current experimental bounds. To illustrate the impact of the different contributions in Fig.~\ref{fig:bsg-tanb} we plot $BR(\bar B\to X_s\gamma)$ with respect to $\tan\beta$ for the three cases where 1) only the charged Higgs contribution is important, 2) only
the chargino-stop loop without $A$-terms is important 3) only the chargino-stop loop with $A$-terms is important.

\section{The SUSY Higgs at the LHC}
For our analysis and plots we use {\tt FeynHiggs}~\cite{feynhiggs} for the Higgs mass, both for the MSSM and the DMSSM,
while we use the one-loop formula from~\cite{Elliott:1993bs} for the NMSSM. Note that there is at least a 2~GeV uncertainty
in the Higgs mass, which is larger in the NMSSM case. Depending on the Higgs mass, we consider $\sigma \times BR$ for different channels normalized to the SM values as described in Tab.~\ref{tab:sBR}.
\begin{table}[t]
\begin{center}
\begin{tabular}{|c|c|}
\hline
\phantom{{}\hspace{4cm} {} }&\phantom{{}\hspace{6cm} {} }\\
$m_h<131.5$ &  $\displaystyle R_{\gamma\gamma}=\frac{\sigma_{gg\to h}}{\sigma_{gg\to h}^{\rm (SM)}}\cdot
	\frac{{\rm BR}_{h\to\gamma\gamma}}{{\rm BR}_{h\to\gamma\gamma}^{\rm (SM)}}$ \\ &\\
\hline &\\
$131.5<m_h<2m_W$ & $\displaystyle  R_{WW^*}=\frac{\sigma_{gg\to h}}{\sigma_{gg\to h}^{\rm (SM)}}\cdot
	\frac{{\rm BR}_{h\to WW^*}}{{\rm BR}_{h\to WW^*}^{\rm (SM)}}$ \\ &\\
\hline &\\
$m_h>2m_W$ & $\displaystyle R_{VV}=\frac{\sigma_{gg\to h}}{\sigma_{gg\to h}^{\rm (SM)}}$ \\ &\\
\hline 
\end{tabular}
\caption{\label{tab:sBR} Higgs cross section times branching ratio normalized to the SM value 
for the different intervals of Higgs masses used in the analysis.}
\end{center}
\end{table}
We use leading order formul\ae~for the widths and cross section~\cite{Kileng},since the leading QCD corrections factorize out in the ratio. In the cases we study, the vector boson fusion channel is always subdominant compared to the gluon-gluon fusion one for the Higgs production cross section and can be safely neglected. 

We also implement LHC bounds using the latest results from Higgs searches with $\sim5$~fb$^{-1}$ \cite{ATLAShiggsresults, CMShiggsresults}.
In particular, for each Higgs mass, we compare the quantities in Tab.~\ref{tab:sBR} with the strongest between the ATLAS and the CMS bounds on the $h\to\gamma\gamma$ channel for $m_h<131.5$~GeV, the CMS bounds on $h\to WW^*$ for $131.5~{\rm GeV}<m_h<2m_W$ and the CMS combination of $h\to WW$ and $h\to ZZ$ for $m_h>2m_W$, respectively. The accuracy of such bounds should be taken with a grain of salt, because of the uncertainties both on the theory side for the Higgs mass calculation and on the experimental side for the energy scaling calibration. For example, a shift of order 1-2~GeV in energy can either amplify or dump sensibly the bounds around 125~GeV where both ATLAS and CMS
have excesses.

Given the sensitivity of $b\to s \gamma$ to $m_A$, $m_{\tilde t_2}$, and $A_t$, we calculate the corresponding bounds separately in each example using the {\tt SUSYBSG} code \cite{SUSYBSG} for light (600 GeV) and heavy (10 TeV) squarks of the first and second family. 

Finally for estimating the fine-tuning we use Eq.~(\ref{fine-tuning}). In the MSSM and DMSSM cases
the main contributions come from the $\mu$-term at tree level and from $m_{\tilde t_i}$ and $A_t$
at one loop. We set the mediation scale to 100~TeV to minimize the tuning. For the NMSSM 
we also consider the contributions coming from the singlet sector.

In all cases, the region with very large stop mixing is disfavored by vacuum instabilities and/or
large corrections to the $\rho$ parameter. Most of this region has been crossed out in the plots
by requiring the Higgs to be above the LEP bounds.

For our calculation we use the following parameters without loss of generality
\begin{center}
\begin{tabular}{|c|c|c|c|c|c|c|}
\hline
$m_t$ & $M_2$ & $M_3$ & $m_{\tilde t_L}^2$-$m_{\tilde t_R}^2$ & $m_{\tilde b_R}$&  $\Lambda$\\ \hline
173.2& 300 & 1000 & $100^2$ & $\infty$ &  $10^5$ \\ 
\hline
\end{tabular}
\end{center}
In particular gauginos masses, whose impact on the quantities we discuss is marginal, are heavy enough not to be ruled out but light enough not to affect the fine-tuning.
The left- and right- hand stop masses are almost degenerate and we have checked that a larger splitting between these soft masses increases only the fine-tuning without altering the conclusion of our analysis. The role of the sbottoms is marginal, and we keep the left-handed sbottom while the other is decoupled.

We follows the conventions in \cite{SLHA2} except for the $\mu$-term, which we define with the opposite sign, and the ordering of the stop eigenvalues, which we choose to be $m_{\tilde t_2}< m_{\tilde t_1}$. For the singlet sector of the NMSSM we instead use the conventions of \cite{Elliott:1993bs} .

\subsection{The Higgs in the MSSM}
\label{MSSMHiggs}
\begin{figure}[t!] %  figure placement: here, top, bottom, or page
 \begin{center}
 \includegraphics[width=6in]{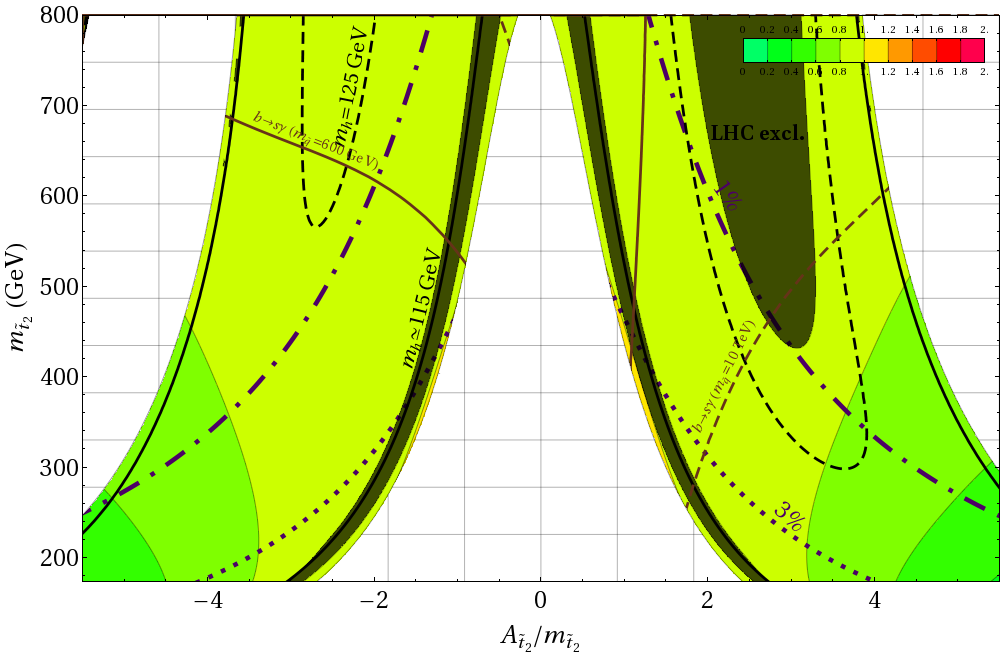}
  \caption{We present Higgs production contours relative to the SM in the MSSM scenario for $m_A=800$ GeV. We block out the region where the Higgs is below the LEP bound. We also draw the contours of $1\%$ (dotted-dashed)and $3\%$ (dotted) tuning. The region allowed by $b\rightarrow s \gamma$ at 95\%~CL lies above the solid (dashed) line when squarks are at 600~GeV (10~TeV). The shaded region is instead excluded by the LHC.}
 \label{MSSMhighmA}
 \end{center}
\end{figure}

\begin{figure}[t!] %  figure placement: here, top, bottom, or page
 \begin{center}
 \includegraphics[width=6in]{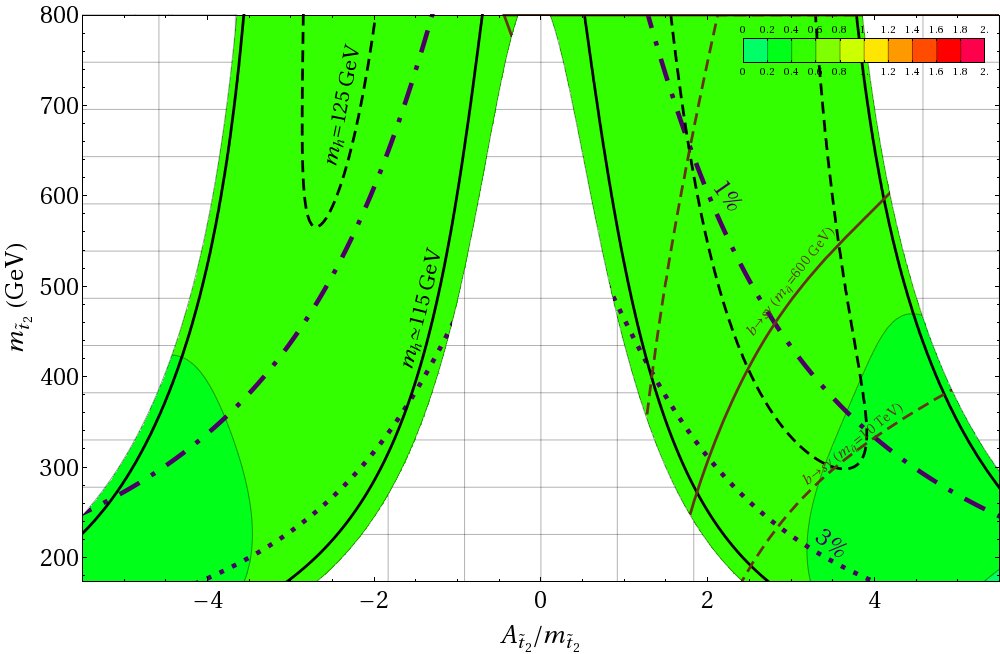}
  \caption{Same as Fig.~\ref{MSSMhighmA}, but with $m_A=250$ GeV.}
 \label{MSSMlowmA}
 \end{center}
\end{figure}

\begin{figure}[t!] %  figure placement: here, top, bottom, or page
 \begin{center}
 \includegraphics[width=6in]{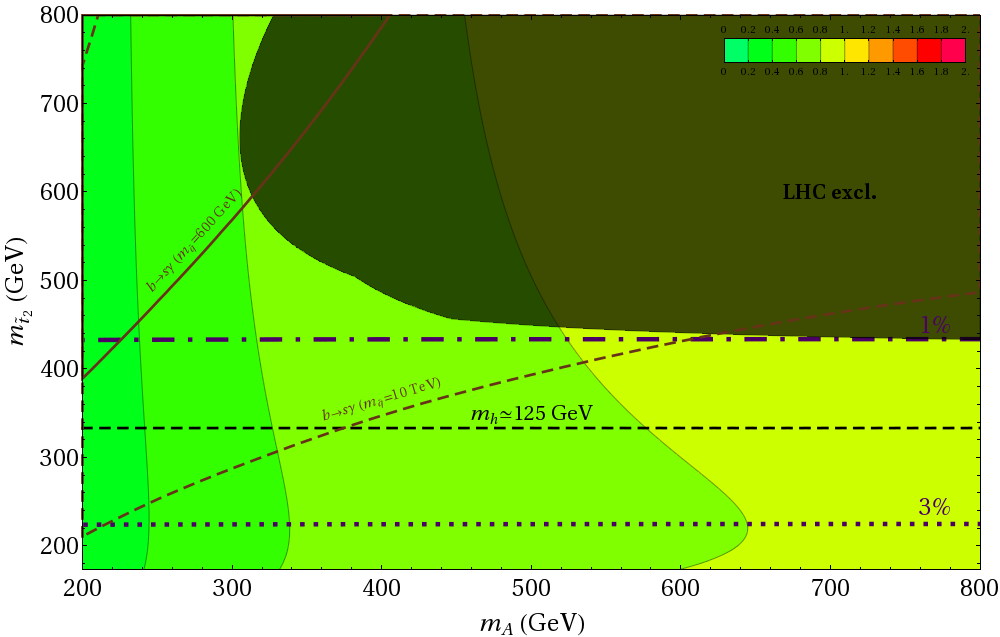}
 \caption{Higgs production relative to the SM as a function of the $m_A$ and the lightest stop mass when $A_t=3 m_{\tilde t_2}$ in the MSSM case. We also present the same Higgs mass, fine-tuning and $b \rightarrow s \gamma$ contours as in Fig.~\ref{MSSMhighmA}.}
 \label{MSSMvaryingmA}
 \end{center}
\end{figure}

In this class of models we consider the minimal SUSY spectrum and we relax the requirement of universality for the scalar masses allowing only the stops to be light. With no additional contributions to the Higgs mass in the MSSM, the radiative corrections from the stop sector are the only way to satisfy the LEP bound. The region where the tuning is minimized and the Higgs is still heavy enough is the one with large stop mixing and $\tan\beta$~\cite{yasunori-Aterms}. In this region of parameter space, the Higgs is light and the most relevant channel  at the LHC is $h\rightarrow \gamma \gamma$, thus
in figs.~\ref{MSSMhighmA},  \ref{MSSMlowmA}, and  \ref{MSSMvaryingmA} only $R_{\gamma\gamma}$ has been used.

For the plots we choose $\mu=200$~GeV, and 
$\tan \beta=10$, to allow for the Higgs to lie above the LEP bounds, but not too large to avoid
stronger bounds from direct and indirect searches. 

In Fig. \ref{MSSMhighmA} we show how $R_{\gamma\gamma}$ varies as a function of the lightest stop mass and the ratio of the $A$-term over the stop mass, when $m_A$=800 GeV. Now, the tree level effects from the Higgs mixing are negligible. We block out the region where the Higgs is below the LEP bound. The Higgs mass increases to $\sim$120($\sim130$)~GeV in the inner part of the allowed region for low(large) values of the stop mass.  The asymmetry of the Higgs mass contours with respect to the stop mixing arises from a two loop gluino-stop correction and is correlated with the relative sign of $A_t$ and the gluino mass\footnote{We thank P. Slavich for clarifying this point.}.
We also show the contours of 1$\%$ and 3$\%$ tuning, which has been estimated through Eq. (\ref{fine-tuning}).
With a large $\tan\beta$ and small $\mu$-term, the main contribution to the tuning comes from the radiative stop corrections, as expected.  Since $m_A$ is large, loop effects are responsible for changing the Higgs production. LEP bounds on the Higgs mass select the region of large $A_t$, where $R_{\gamma\gamma}$ gets suppressed. The suppression is however only large for very low stop masses, which are incompatible with $b\to s \gamma$ bounds. The latter push us to the decoupling limit, where the Higgs looks more SM-like, with a $\sim20\%$ reduction of its production cross-section at best. Low fine-tuning though in this case still favors the presence of one stop at around 400~GeV, which could be within the reach of the 7~TeV LHC. 
Finally we can also see how the LHC started probing this low Higgs mass region: masses around 115~GeV have just been  ruled out by ATLAS, while masses above 127~GeV from CMS.


In Fig.~\ref{MSSMlowmA} $m_A$ is now lowered to 250~GeV, Higgs mixing effects become important and BR($h\to \gamma \gamma$) is suppressed compared to its SM value because $h \rightarrow b \bar b$ is enhanced. At the same time, pushing the Higgs mass above the LEP bound requires large $\frac{A_t}{m_{\tilde t_2}}$, further suppressing $R_{\gamma\gamma}$, which is now roughly around half the SM value in most of the allowed parameter space. In this case, the Higgs remains hidden at the LHC, probably till the end of next year. It is worth pointing out, however, that the requirement of small tuning favors the presence of a light stop at around 300 GeV which may show up before the Higgs is discovered.

In Fig.~\ref{MSSMvaryingmA}, keeping $A_t=3 m_{\tilde t_2}$, we show how  the shape of the contours changes by increasing $m_A$ signaling the decoupling of the tree-level effects from Higgs mixing.  Higgs mixing effects can easily provide a $40\%$ suppression of $gg\rightarrow h \rightarrow \gamma \gamma$ compared to the SM, but they become unimportant for $m_A>400$ GeV as expected, since we are well within the decoupling limit of the Higgs sector. Fig.~\ref{MSSMvaryingmA} also shows that the fine-tuning is insensitive to $m_A<1$~TeV because its contribution to the fine-tuning is suppressed by $\tan\beta$, as discussed previously. $b\to s \gamma$ bounds push towards heavier stop masses (and Higgs masses as well), unless $m_A$ is light enough to allow for a cancellation between the charged Higgs and the stop-chargino loops.

In conclusion we see that the Higgs is always light in the minimal implementation of supersymmetry. 
In the less tuned region, the requirement of large $A_t$ in order to push the Higgs mass above the LEP bound guaranties that $R_{\gamma\gamma}$ will always be smaller than one. This suppression is enhanced further when the pseudoscalar Higgs is light and the Higgs can now be hidden from the LHC till the end of next year.

\subsection{The Higgs in the MSSM with extra D-terms}
\label{DtermHiggs}

\begin{figure}[t!] %  figure placement: here, top, bottom, or page
 \begin{center}
 \includegraphics[width=6in]{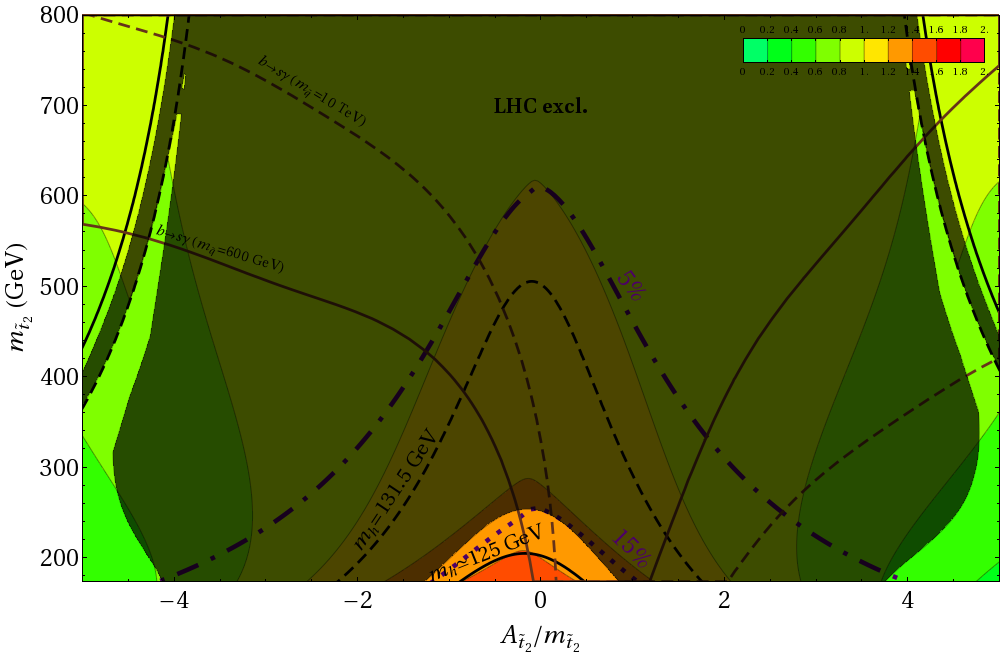}
 \caption{Higgs production at the LHC compared to the SM when Higgs D-terms are added to the MSSM with $m_A=800$~GeV in the $A_t/m_{\tilde t_2}-m_{\tilde t_2}$ plane. We show the contours of $15\%$ (dotted) and $5\%$ (doted-dashed) fine-tuning as well as the Higgs mass contours of 125~GeV and 131.5~GeV.We also show the regions allowed by $b \rightarrow s \gamma$ at 95\% CL and the shaded region ruled out by the LHC, as in the case of the MSSM.}
 \label{QMSSMhighmA}
 \end{center}
\end{figure}

\begin{figure}[t!] %  figure placement: here, top, bottom, or page
 \begin{center}
 \includegraphics[width=6in]{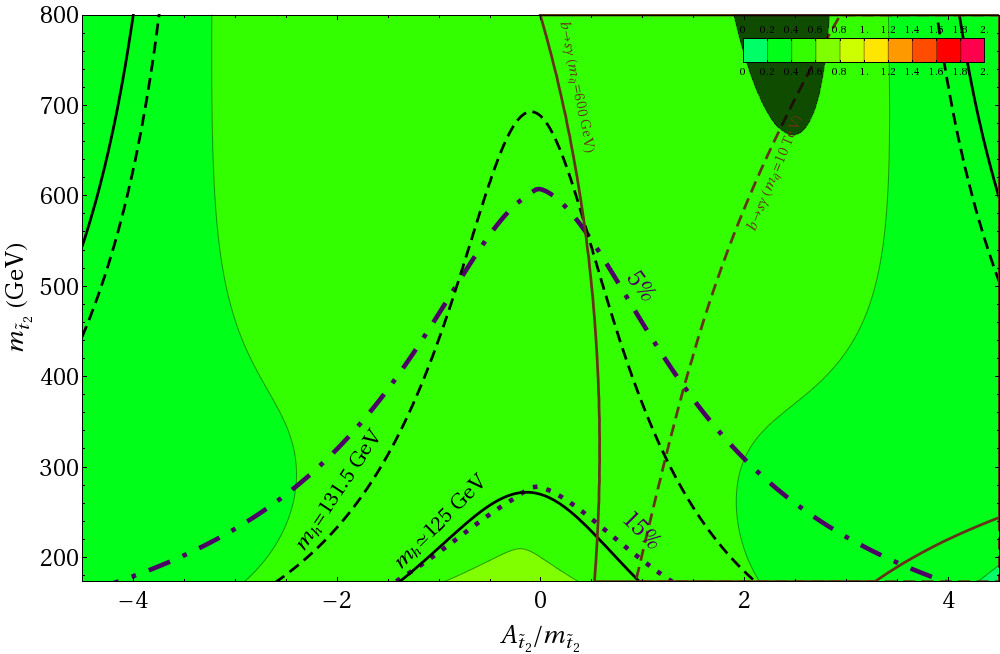}
 \caption{Same as Fig.~\ref{QMSSMhighmA} but with $m_A=250$~GeV.}
 \label{QMSSMlowmA}
 \end{center}
\end{figure}

\begin{figure}[t] %  figure placement: here, top, bottom, or page
 \begin{center}
 \includegraphics[width=6in]{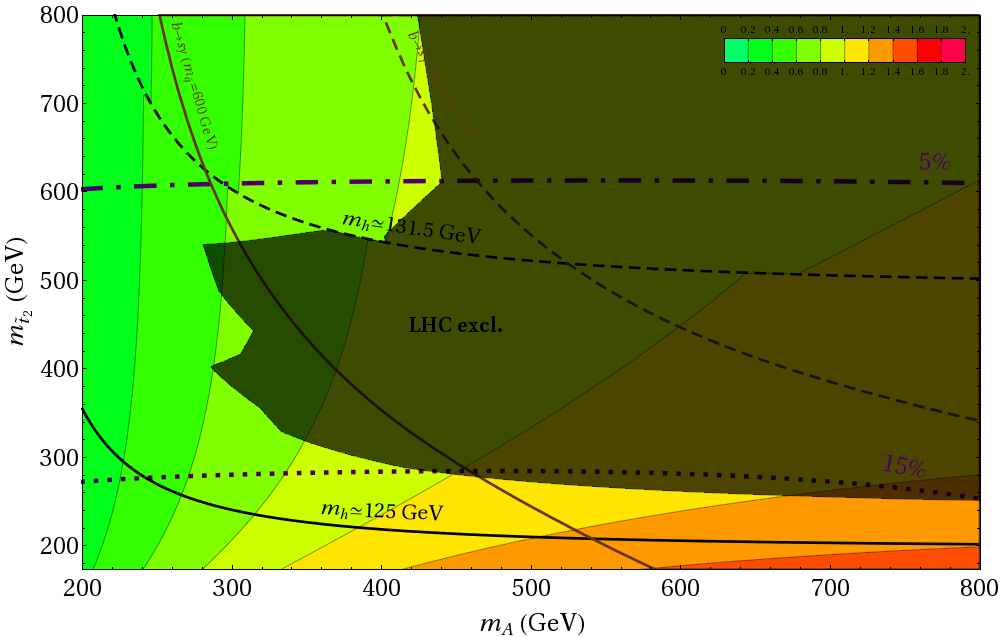}
 \caption{Higgs production relative to the SM as a function of the $m_A$ and the lightest stop mass when $A_t=0$ in the MSSM with D-terms. We also present the same Higgs mass, $b \rightarrow s \gamma$ and fine-tuning contours as in Fig. \ref{QMSSMhighmA}.}
 \label{QMSSMvaryingmA}
 \end{center}
\end{figure}

In these models, the SM gauge sector is extended with an extra gauge group that is broken not much above the weak scale and under which the Higgs is charged \cite{D-terms}. When the field responsible for the breaking of that group gets a soft mass around the weak scale, the D-term for the Higgs results in an increase of the Higgs quartic and the tree level mass of the Higgs gets an additional contribution. We explore this scenario assuming that the D-term is generated by a $U(1)_x$ symmetry under which $h_u$ and $h_d$ are vector-like and the new contribution to the Higgs potential is given by
\bea
\delta V=\frac{\lambda_x^2}{8} (h_u^2-h_d^2)^2\,.
\eea
We choose $\lambda_x=0.7$, large enough to allow the Higgs mass to lie confortably above the LEP bounds without the need of $A$-terms but low enough such that the corresponding $U(1)_x$ can satisfy electroweak precision tests and collider searches \cite{Lodone:D-terms}. We keep the rest of the parameters for the charginos and the sbottoms the same as in the MSSM case (see section \ref{MSSMHiggs}) but change $\tan \beta$ to 5. We now find that the Higgs easily is above the LEP bound in the entire region of the parameter space we are considering. According to Tab.~\ref{tab:sBR}, since the Higgs mass never exceeds $2m_W$, in Fig.~\ref{QMSSMhighmA}, \ref{QMSSMlowmA}, and \ref{QMSSMvaryingmA}, we show $R_{\gamma\gamma}$ when the Higgs is between the LEP bound and 131.5 GeV, and $R_{WW^{*}}$ when the Higgs is above 131.5 GeV.

In Fig. \ref{QMSSMhighmA} we show $R_{\gamma\gamma}$ and $R_{WW^*}$ as a function of the lightest stop mass and the ratio $\frac{A_t}{m_{\tilde t_2}}$ for $m_A$=800 GeV. The fine-tuning is significantly relaxed due to the extra tree-level contribution to the Higgs mass and we now plot the contours of $5\%$ and $15\%$ fine-tuning. The lightest Higgs state can now be heavier, and loop effects become more pronounced in the decoupling limit of the heavy Higgs sector. Since the region $A_t\sim0$ is not excluded anymore, there is the possibility of $R_{\gamma\gamma}>1$ provided that bounds from $b\rightarrow s \gamma$ are satisfied. This is possible when the rest of the squarks are light ($\sim$600~GeV), otherwise the Higgs is pushed to a region when it looks more SM-like. In this case, $b \rightarrow s \gamma$ seems to constrain the hierarchy between the third and the other two families in split family scenarios when a gauge mediation mechanism is present and the $A$-terms are small. When we apply the latest LHC Higgs search results, we see that most of the parameter space is now excluded leaving only open the possibility  of very light stops and small $A$-terms where the Higgs production cross-section  is enhanced up to $50\%$. Note that this region is still allowed in the plot because of the ATLAS and CMS excesses around 125~GeV.

In Fig.~\ref{QMSSMlowmA}, we present the same parameter space when $m_A$=250 GeV.  Similarly to the MSSM case, the effects of the light pseudoscalar mass are evident everywhere in the parameter space, due to the increase of the $h\rightarrow b \bar b$ width, but now the suppression can be smaller as small $A$-terms are allowed. Rare B decays now exclude the possibility of zero $A$-terms and further constrain split family scenarios when a gauge mediation mechanism is considered.

Finally in Fig.~\ref{QMSSMvaryingmA} we show explicitly the dependence on $m_A$,  keeping fixed $A_t=0$, which is
the value preferred by low scale SUSY-breaking scenarios, such as gauge (GMSB) and anomaly mediation (AMSB). The slight difference between $R_{\gamma\gamma}$ and $R_{WW^*}$ across the 131.5~GeV Higgs mass contour is due to Higgs mixing effects suppressing the Higgs coupling to $WW$ less than the Higgs coupling to tops. The overall width $h\rightarrow \gamma \gamma$ gets a small boost in this region because the top contribution now subtracts less from the $W$ contribution and $R_{\gamma\gamma}$ tends to be a bit higher than $R_{WW^*}$.
$b\to s \gamma$ bounds push either the stop or $m_A$ to be heavy, with a preference for light 1st and 2nd generation squarks.
After LHC bounds are taken into account, only two regions are singled out, one more tuned with large stop masses 
and reduced coupling of the Higgs (due to Higgs mixing) and one with light stops, heavy $m_A$ and enhanced $R_{\gamma\gamma}$. A SM-like Higgs in this scenario is very unlikely.

In conclusion, in this kind of models generically the Higgs cross section and branching ratios differ from the SM even
by order one. Strong bounds from the LHC apply and when combined with bounds from $b\to s \gamma$ and naturalness
they strongly constrain the allowed parameter space, especially in scenarios with vanishing squark mixing.

\subsection{The Higgs in the NMSSM}
\label{NMSSMHiggs}

Motivated by the $\mu$ problem, the NMSSM extends the Higgs sector by a singlet that now mixes with the MSSM Higgses. We consider the $Z_3$ symmetric version of the NMSSM with the following superpotential:
\bea
W=\lambda S H_u H_d - \frac{\kappa}{3} S^3
\eea
Symmetry breaking in the singlet sector strongly depends on the singlet $A$-terms, $\lambda A_\lambda S H_u H_d$ and $\kappa A_\kappa S^3/3$, which have to be non-zero for the singlet to acquire a vev. There are also requirements on these terms from vacuum stability \cite{NMSSMreview, arXiv:1005.1070}.

\begin{table}[t]
\end{table}

\begin{figure}[t!] %  figure placement: here, top, bottom, or page
 \begin{center}
 \includegraphics[width=6in]{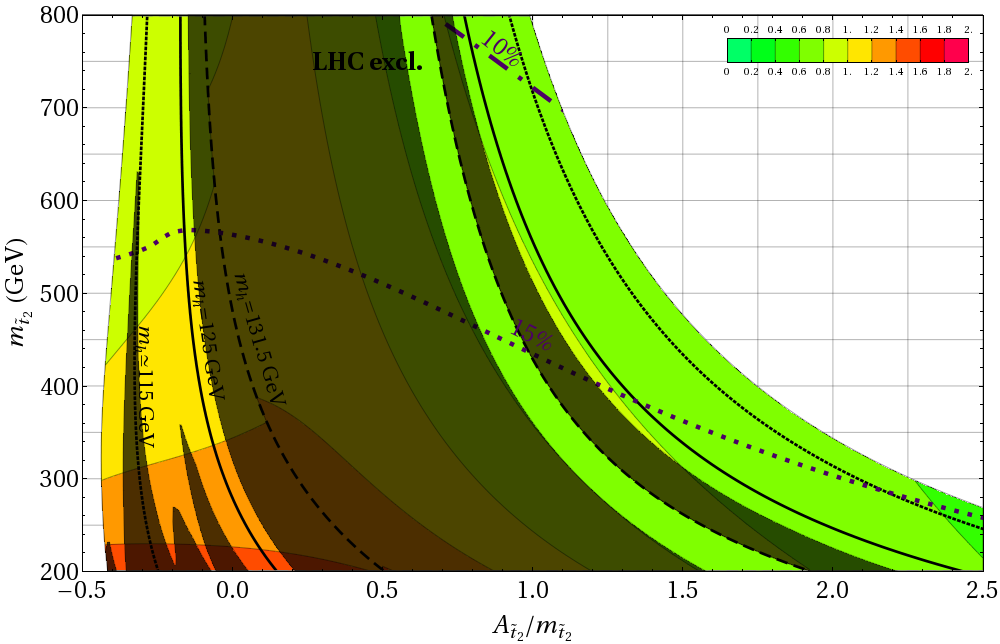}
 \caption{We present the contours of Higgs production relative to the SM in the  perturbative NMSSM scenario. We draw the contours of $10\%$ and $15\%$ tuning. We also show the 115, 125 and 131.5 GeV Higgs mass contours. The dark shaded area has been excluded by LHC Higgs searches.}
 \label{NMSSMsmalllambda}
 \end{center}
\end{figure}

\begin{figure}[t] %  figure placement: here, top, bottom, or page
 \begin{center}
 \includegraphics[width=6in]{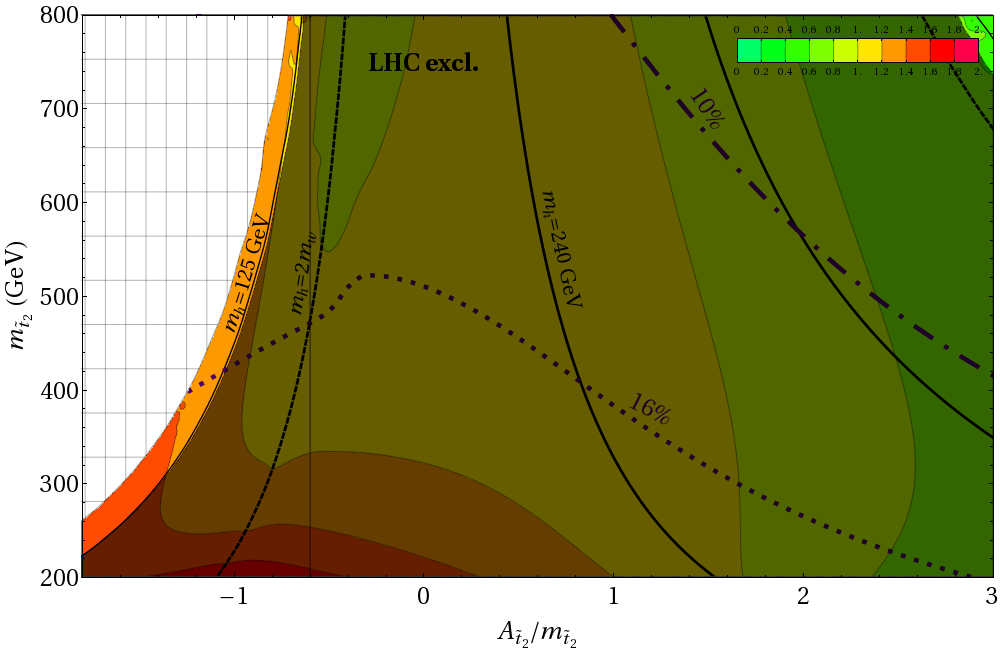}
 \caption{We present the contours of $\sigma \times BR(h \rightarrow WW^{(*)})$ relative to the SM in the $\lambda{SUSY}$ scenario. We draw the contours of $10\%$ and $16\%$ tuning. We also show the 125, $2m_W$, and 240~GeV Higgs mass contours. The dark shaded area has been excluded by LHC Higgs searches.}
 \label{NMSSMlargelambda}
 \end{center}
\end{figure}

We select two qualitatively different regions of the NMSSM parameter space (in GeV)
\begin{center}
\begin{tabular}{|c|c|c|c|c|c|c|c|c|c|c|}
\hline
$\mu=\lambda v_s$ & $\tan\beta$ &   $\lambda$&$\kappa$&$A_\lambda$& $A_{\kappa}$\\ \hline
200  & 1.5 &0.7&-0.7& 750 & -750 \\ 
\hline
300 & 1.4 & 1.4&-1.4& 1000 & -1000\\ 
\hline
\end{tabular}
\end{center}
The first we refer to as perturbative NMSSM, see Fig.~\ref{NMSSMsmalllambda}, the second as $\lambda{SUSY}$, see Fig.~\ref{NMSSMlargelambda}. Both regions are chosen such that there are no pseudoscalar Higgs states for the light scalar Higgs to decay into that can drastically alter the Higgs width. Indirect bounds from $b \rightarrow s \gamma$ are now irrelevant since at small $\tan\beta$ the SUSY contributions  are smaller and they always tend to cancel (see Fig.~\ref{fig:bsg-tanb}).

In the perturbative NMSSM, we choose small values of the singlet couplings that can remain perturbative up to the GUT scale. The value of $\tan\beta$ now is small in order to maximize the contributions to the Higgs mass. The Higgs mass takes values up to 150~GeV. In Fig.~\ref{NMSSMsmalllambda}, we present $R_{\gamma\gamma}$ and $R_{WW^*}$ as a function of $m_{\tilde t_2}$ and $A_t/m_{\tilde t_2}$. We can easily see that mixing effects in the Higgs sector are very suppressed as $\tan\beta$ is small and $h\rightarrow b\bar b$ does not change a lot compared to its SM value. In addition, the preferred $A$-terms are small and the Higgs starts being overproduced in a large part of the parameter space; the Higgs is slightly heavier than the previous two cases and loop effects decouple more slowly with increasing stop mass. In fact, current LHC searches highly constrain a big part of this parameter space, leaving out the regions
with $m_h\lesssim130$~GeV, where the bounds are weaker. In particular the region near $A_t\sim0$ is still allowed despite the enhancement because of the excess around 125~GeV in the ATLAS and CMS results. The other region which survives is the one with very large $A$-terms where the Higgs production cross section is sufficiently suppressed. Most of these regions of the parameter space should be covered by the end of next year.

In $\lambda{SUSY}$ \cite{Harnik:2003rs, lambdaSUSY}, the perturbativity condition at $M_{GUT}$ on the couplings is relaxed, 
with virtually no bounds on the value of the Higgs mass. For the parameters we have chosen,
the Higgs becomes as heavy as $\sim$250~GeV. In most of the parameter space $m_h>2m_W$  and we only plot $\sigma \times BR(h \rightarrow WW^{(*)})$. The mixing in the Higgs sector is not as important because the Higgs decay to $b \bar b$ no longer dominates the Higgs width. Loop effects become more important for heavier stops due to the large Higgs mass, and the Higgs is overproduced for a large part of the parameter space, when the $A$-terms are not too large~\cite{arXiv:0710.5750}. The strength of the LHC bounds for a heavy Higgs together
with this enhancement of the cross section make most of the parameter space excluded. The only region still allowed 
is the one with small Higgs mass, near the boundary of stability, where there is an enhancement due to large mixings with the heavier Higgs states.
We will return on this particular region in section~\ref{Higgsmixing}. In this region we still plot $\sigma \times BR(h \rightarrow WW^{*})$
because this channel is enhanced with respect to the diphoton one in this model and, even though the LHC bounds from $h\to\gamma\gamma$ 
are stronger than those from $h\to WW^*$ for a SM Higgs, the former are not constraining in this case.

It is fair to say that $\lambda{SUSY}$ is highly disfavored by LHC searches, although corners of parameter space are still allowed, as well as the possibility to hide the Higgs through decay to light pseudoscalars.

\subsection{Large Mixing between Higgs States}
\label{Higgsmixing}

\subsubsection{Mixing in the MSSM}
We discuss now the case of large mixing between the light and the heavy Higgses in the MSSM, corresponding to
values of $m_A\gtrsim 100$~GeV and $\tan\beta\lesssim10$, still allowed experimentally (see Fig.~\ref{fig:tanbma}).
In this special region, the light and the heavy Higgs do not have a  large mass splitting and the role of the SM Higgs is almost equally shared between the two states. The behavior of the Higgses strongly depends on their mass. As already discussed at the beginning of section~\ref{SUSYHiggs}, while the couplings to the top and the $W$ are both suppressed compared to their SM values, the coupling to the bottom is enhanced. The production of both Higgs states at the LHC can be highly suppressed, when the $h\rightarrow \gamma \gamma$ and $h\rightarrow W W^{*}$ channels are relevant. At large $\tan\beta$, the pseudoscalar Higgs may signal the presence of an extended Higgs sector at the LHC, since its couplings are not affected by the mixing and are enhanced by $\tan \beta$. When the width is no longer dominated by decays to $b \bar b$, we expect one Higgs to be visible, when the cross-section of the other Higgs is suppressed, given that
\bea
y_{t \bar t h}^2+y_{t \bar t H}^2= \frac{1}{\sin^2\beta}\,,
\eea
where $y_{t \bar t h}$ and $y_{t \bar t H}$ are the top couplings to the light and heavy Higgs, respectively. 

\begin{figure}[t!]
\begin{center}
(a)\hspace{15cm} (b)  \\
\includegraphics[height=0.3\textwidth]{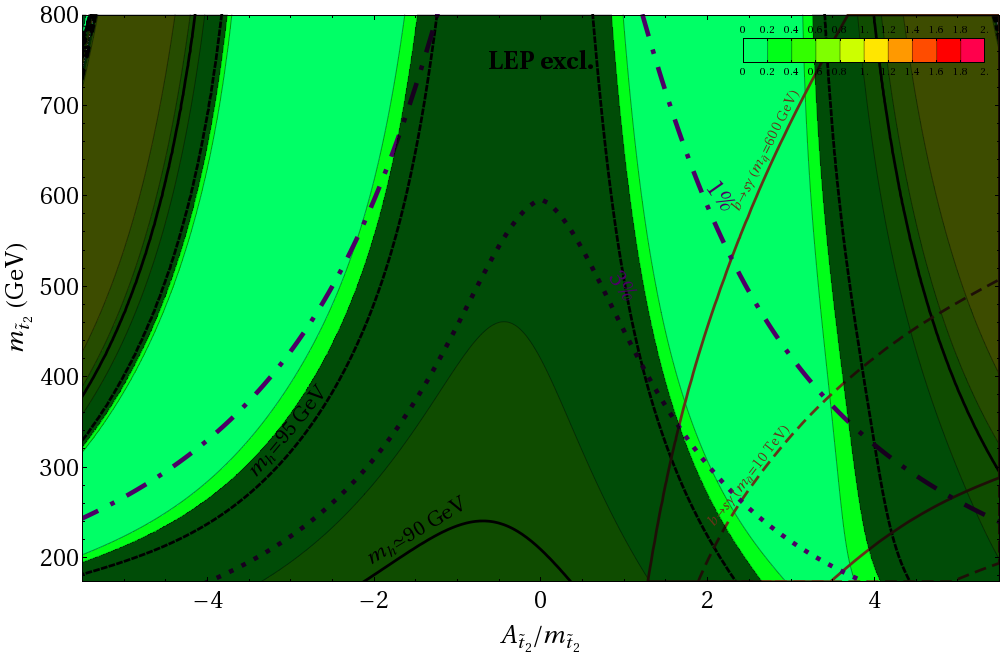} 
\includegraphics[height=0.3\textwidth]{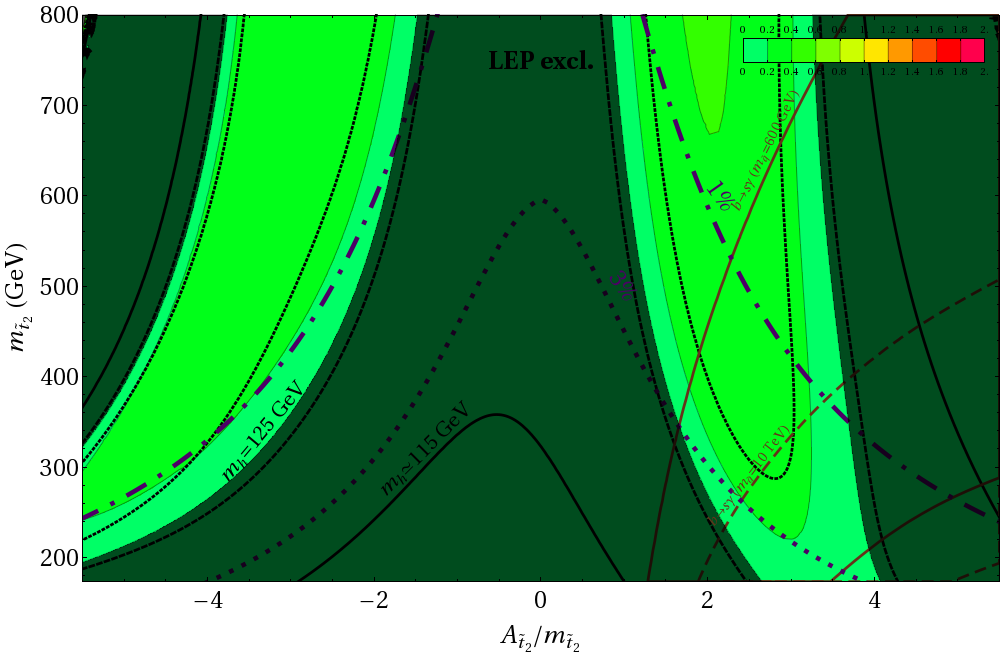} \\
%(c)\hspace{15cm} (d)  \\
%\includegraphics[height=0.3\textwidth]{HiggsCS-QMSSM4l.png} 
%\includegraphics[height=0.3\textwidth]{HiggsCS-QMSSM4h.png} \\
%(e)\hspace{15cm} (f)  \\
%\includegraphics[height=0.3\textwidth]{HiggsCS-NMSSM2a-shaded.png} 
%\includegraphics[height=0.3\textwidth]{HiggsCS-NMSSM2b-shaded.png} 
\caption{Contours of Higgs production relative to the SM when the Higgs maximally mix in the MSSM. We show tuning contours (dotted and dashed-dotted), $b\rightarrow s \gamma$ allowed regions between the solid (dashed) for light (heavy) squarks, and different Higgs mass contours. Finally, the dark shaded areas are excluded by Higgs searches.}
\label{mixing}
\end{center}
\end{figure}

In Fig.~\ref{mixing}(a) and (b), we show an example for the light and the heavy Higgs contours of $R_{\gamma\gamma}$  in the MSSM scenario. 
The parameters used are the same as in section~\ref{MSSMHiggs} except for $\tan\beta$ which we set to 8 and $m_A=100$~GeV.

When the Higgs mass is below $\sim$114~GeV, LEP searches are the most sensitive and we substitute $R_{\gamma\gamma}$ for the ratio of the Higgs coupling to the $W$
\bea
&R^h_{WW^*}=\sin^2(\beta-\alpha)\,,\\
&R^H_{WW^*}=\cos^2(\beta-\alpha)\,.
\eea

We also show the excluded region from LEP bounds on the light Higgs, which significantly constrain the parameter space for the heavy Higgs as well. For the parameter space still not probed, the increase in $h\rightarrow b\bar b$ for both Higgses hides them at the LHC, but the pseudoscalar Higgs is light enough and it is well within the reach of the next round of $A\rightarrow \tau^+ \tau^-$ searches at the LHC, as can be seen in Fig.~\ref{fig:tanbma}. 

The scenario with additional D-terms for the Higgs differs in that the heavy Higgs can be heavy enough for decays to $b \bar b$ to be irrelevant and its LHC signals are enhanced. As a result, a lot of the parameter space in this case can be constrained from the absence of the heavy Higgs.

\subsubsection{Mixing in the NMSSM}

In the NMSSM, the addition of the singlet affects the Higgs couplings to the SM particles in two ways. First, it suppresses all Higgs couplings by directly mixing with $H_u$ and $H_d$. Another effect comes from the $SH_uH_d$ coupling contribution to the Higgs quartic in the F-term potential. Analogously to the DMSSM case, the effects of this quartic do not decouple when the singlet gets its mass from SUSY breaking. This new contribution is always positive contrary to the MSSM where it is generated by D-terms\footnote{We thank R.~Rattazzi for pointing this out.}. The top (bottom) Yukawa is now enhanced (suppressed) compared to its SM value, opposite to what is happening in the MSSM as seen in Eq. (\ref{yukawasMSSM}). For a light Higgs, this leads to an enhancement of $R_{WW^*}$  because the Higgs width to bottoms decreases.

These two effects mentioned above combine in the left side of Fig.~\ref{NMSSMlargelambda} to enhance the Higgs $\sigma \times  BR$. In this region, even though the singlet and the heavy Higgs are always above 400 GeV there are still large mixing effects because the lightness of the Higgs is driven by cancellations in the $3\times3$ Higgs mass matrix. For the same reason, the heavy Higgs, which is around 400 GeV, has non negligible couplings to vector bosons and now LHC searches for a heavy SM Higgs may place the strongest bound on this region.

\subsection{Higgs coupling determination at the LHC}

It is obvious that deviations of the Higgs from the SM behavior may signal the presence of new light degrees of freedom. But how well can the LHC measure these deviations after the Higgs is discovered? QCD uncertainties in the Higgs production  introduce $10-20\%$ error in the measurement of $\sigma \times BR$ for every Higgs decay channel. 
In most cases, absolute coupling determination cannot be better than $20\%$ \cite{higgscouplingsatlhc1, higgscouplingsatlhc2}.  These uncertainties though disappear when considering the ratios of $\sigma \times BR$ for channels that have the same production mechanism. In this case, most systematics are of the order of a few $\%$ and it is possible for the LHC to become a precision probe of Higgs couplings. An example of such a ratio for a Higgs below $\sim$130 GeV is:
\bea
\frac{\sigma \times BR(h \rightarrow \gamma \gamma)}{\sigma \times BR(h \rightarrow Z Z^{*})} \,.
\eea 
For both channels, energy resolution is excellent and LHC will have enough statistics to eventually be systematics limited. This ratio in the MSSM becomes sensitive to stop loop effects which can be as large as $50\%$ in the large mixing region. For example, an accuracy of 3\% in such ratio may probe stop masses as high as 500~GeV. The ratio becomes sensitive to mixing effects as well, when $\tan \beta$ is small. This ratio can be enhanced compared to the SM value only when stop mixing is large or the top coupling is suppressed by mixing effects. In this case, the Higgs production cross-section will also be suppressed.

Another possible ratio is the photon channel compared to the $\tau^+ \tau^-$ (or $b\bar b$) channel. This ratio would be simultaneously sensitive to mixing and loop effects, but it is also harder to measure since it relies on vector boson fusion and suffers from low statistics. Note that an early analysis of coupling ratios was performed in~\cite{higgscouplingsatlhc2}, but it was limited by the amount of data used in the analysis and by choosing the $h\rightarrow WW^*$ the main channel to which all channels are compared. Of course, any systematics that affect the efficiency of signal identification will make the sensitivity of these ratios worse; a full analysis is required in order to properly estimate the final precision with which these coupling ratios can be measured.

\subsection{The possibility of a 125~GeV Higgs}
\label{Higgsat125}

The recent ATLAS search for the SM Higgs with 5~fb$^{-1}$ shows an almost 3$\sigma$ excess in the diphoton and 4-lepton channels at around 125~GeV \cite{ATLAShiggsresults}, and CMS has also a less significant excess in the diphoton channel at a nearby energy \cite{CMShiggsresults}. In this section we would like to contemplate  what a $\sim$125~GeV Higgs would imply for Supersymmetry and the possibility of sparticles light enough to be discovered at the LHC, should such an excess turn into a discovery. 

For the MSSM, as can be seen in Section \ref{MSSMHiggs}, a Higgs mass at 125~GeV can still be achieved with light stops and large enough $A$-terms. Nevertheless the parameter space becomes quite constrained, once the relatively large production cross-section that is needed to fit the current excess, as well as the bounds from $b\rightarrow s \gamma$ are taken into account. As can be seen comparing Fig. \ref{MSSMhighmA} and \ref{MSSMlowmA}, the mixing effects from $m_A$ have to be decoupled and it is pushed to be heavier than 800~GeV. At the same time, stops have to be heavier than at least 500~GeV making the Higgs look very SM-like---the possibility of a natural SUSY spectrum in this case becomes more remote.

The situation changes drastically once we consider simple extensions of the MSSM where the Higgs gets new tree level mass contributions, such as the DMSSM and the NMSSM. In both these cases, large $A$-terms and heavy stops are no longer necessary to raise the Higgs mass at 125~GeV. Stops can be easily within the reach of the 7~TeV LHC and enhance $gg\rightarrow h \rightarrow \gamma \gamma$ by $50\%$. This enhancement is compatible with the current excess in \cite{ATLAShiggsresults,CMShiggsresults}. 

\section{Conclusions}

The scenarios presented above provide a qualitative picture of the light SUSY Higgs production and decay at the LHC. The discovery of a Higgs with enhanced $\sigma \times BR$ implies that the stops are light and the $A$-terms are small by interfering constructively with the top in gluon-gluon fusion. In particular for GMSB/AMSB scenarios, current LHC and $b\rightarrow s \gamma$ bounds in combination with naturalness already favor this possibility. The mixing of $H_d$ with a singlet in the NMSSM also leads to an enhancement of $\sigma \times BR$ for a light mass Higgs, because it suppresses $h\rightarrow b \bar b$.

Conversely, the discovery of a Higgs with reduced $\sigma \times BR$ compared to its SM value may suggest that the $A$-terms are large, since stops now interfere destructively with the top in $gg \rightarrow h$. $BR(h\rightarrow \gamma \gamma (\text{or}~ WW^*))$ is also suppressed for a light Higgs when mixing between $H_u$ and $H_d$ increases the Higgs width to $b \bar b$.  Finally, when a singlet is added to the Higgs sector and $\tan \beta$ is small, $H_u$ mixing with the singlet decreases the coupling to the top and gluon-gluon fusion is suppressed.

The size of the above effects depends on the ratio of scales between the new physics and the Higgs. The Higgs appears SM-like when the new states are pushed a few times above the Higgs mass. If a non-SM Higgs is detected at the LHC, measuring its mass and how much its couplings deviate from the SM could point to the scale of new light degrees of freedom. So, it is important to further study whether taking the ratios of different channels can improve the accuracy of Higgs coupling measurements at the LHC. 

Finally, if the $\sim125$~GeV excess observed by ATLAS persists, it can be accommodated in the MSSM with large A-terms but, when $b \rightarrow s \gamma$ bounds are taken into account, stops are pushed to be above 500~GeV and the Higgs appears SM-like. In the MSSM, the production of a 125~GeV Higgs is always suppressed compared to its SM value. Simple extensions of the MSSM allow for lighter stops that can now enhance the Higgs production cross-section. This enhancement may be favored by data and points to the presence of light stops in the SUSY spectrum. In this case, a 125 GeV Higgs could be our first evidence of naturalness in SUSY.

\section*{Acknowledgements}

We would like to thank Savas Dimopoulos, Diego Guadagnoli, and Michele Papucci for useful discussions. This work was partially supported by ERC grant BSMOXFORD no. 228169.

\end{document}